\documentclass[twocolumn,twoside,superscriptaddress,prl,aps, showpacs]{revtex4}
\usepackage{amsmath,mathrsfs,amsbsy,color,graphicx,bm,amsthm,amsfonts}
\usepackage{units}
\usepackage{bbm}
\usepackage{bm}
\usepackage{graphics}

\begin{document}

\title{Operation triggered quantum clock synchronization}
\author{Jie-Dong Yue}
\affiliation{Beijing National Laboratory for Condensed Matter Physics, Institute of
Physics, Chinese Academy of Sciences, Beijing 100190, China}
\author{Yu-Ran Zhang}
\affiliation{Beijing National Laboratory for Condensed Matter Physics, Institute of
Physics, Chinese Academy of Sciences, Beijing 100190, China}
\author{Heng Fan}
\email{hfan@iphy.ac.cn}
\affiliation{Beijing National Laboratory for Condensed Matter Physics, Institute of
Physics, Chinese Academy of Sciences, Beijing 100190, China}

\begin{abstract}
We present a novel quantum clock synchronization(QCS) scheme of multiple parties which uses operation as the trigger to start the evolution of the initial state, where existing QCS protocols use measurement to start the evolution. Since the trigger is the unitary operation, we have protected entanglement of remote nodes, and after concentration of the qubits to the center node, general measurements of the total state is possible. We show that our protocol links the QCS problem to a multiple phase estimation problem. We use the Fisher information to give the precision of the synchronization, and explicitly show that the Heisenberg scale of synchronization is achieved in the two party case. We also show that our protocol is very efficient in synchronizing a clock to the average time of other clocks. The precision has an $O(\sqrt{d})$ advantage of the precision of the average time's estimation over the best possible strategy using measurement triggered QCS. The precision is also Heisenberg scale.
\end{abstract}

\maketitle


\textit{Introduction.}---
The synchronization of clocks with high precision is required in many modern technologies and researches, such as navigation, distributed computation, long baseline interferometry, tests of theory of general relativity, laser interferometer gravitational wave observation(LIGO), and telecommunication. Two standard methods of clock synchronization are Einstein's synchronization procedure \cite{einstein1905}, which uses an operational line-of-sight exchange of light pusles between two spatially separated clocks and Eddington's slow clock transport \cite{Eddington1924}, which is based on the internal time evolution of quantum systems. A quantum algorithm for distributed clock synchronization has been proposed by Chuang \cite{Chuang2000}.

A method of quantum clock synchronization(QCS), which uses shared prior entanglement resources to synchronize clocks of two parties, is proposed by Jozsa et al \cite{Josza2000}. The accuracy of the QCS protocol is independent of the participants' relative locations and the properties of the intervening media, which is a major difference and advantage over the standard Einstein's and Eddington's procedures. The QCS protocol has been generalized to the multiparty case \cite{Krco2002}, and the synchronization precision using different initial states for the QCS protocol is also studied \cite{Ben2011, Ren2012, YLZhang2013}. Optimization and limiting issues in the quantum clock literature have been dealt with, with respect to QCS\cite{Buzek1999, Preskill, Yurtsever2002, Giovannetti2001}. Experimental work has also been done on QCS implementation. A distant clock synchronization(picosecond resolution at 3 km distance) based on entangled photon pairs is reported\cite{Valencia2004}.

Applying the technique of quantum metrology, it has been shown that the standard quantum limit $1/\sqrt{N} $ of the parameter estimation precision, can be improved to the Heisenberg limit $1/N$, with entanglement resources employed, where $N$ is the number of particles used in the measurement\cite{Wineland1992, Wineland1994, Giovannetti2004, Giovannetti2006}. In the study of quantum metrology, quantum Fisher information and quantum Cram\'{e}r-Rao bound provides a basic approach\cite{Cramer, Paris2009}. The enhancement of the estimation precision to the Heisenberg limit is the main concern of quantum metrology, and a lot work has been done, both theoretically and experimentally\cite{Giovannetti2004, Giovannetti2006}.

In this letter, we relate the quantum clock synchronization protocol of $d$ clocks to a standard multiple phase estimation problem. We will explicitly show that in the case $d=2$, our procedure gives Heisenberg limit of estimating the time difference. In previous proposed protocols of QCS\cite{Krco2002, Ben2011, Ren2012, YLZhang2013}, the trigger of evolution is the measurement, after which the entanglement will be broken, and the measurement results at different nodes are the useful data which can be used to calculate the time differences, we mention that here only local measurement at different nodes is allowed. For our protocol, unitary operations at different nodes serve as the triggers of evolution of the states, and after evolution states at different nodes will be transferred to the center node through quantum channels, where a general measurement of the total state will be made. We show that our procedure is advantageous when only the average time is needed, that is, when we need synchronize a clock $0$ to the average of $d$ clocks by calculating $\bar{\theta} = t_0 - \Sigma_{i= 1}^d t_i/d$. This is meaningful when instead of synchronizing the clocks of satellites at various places to set up a standard time, we can just take the average of these clocks as the standard time. $\delta \bar{\theta}$, which is the standard deviation of $\bar{\theta}$, will have a $O(\sqrt{d})$ advantage in our proposal compared with when only local measurement is allowed. We should note that in our meaning of clock synchronization, we do not consider the frequency variance of the clocks, what we are concerned is that the clocks may not agree on a common time at the same readout, and we are aiming at synchronizing the clocks to a common time. A scheme of a quantum network of clocks has been proposed, which combines precision metrology and quantum networks to achieve greater clock stability\cite{Komar2014, Kessler2014}.

\textit{Measurement triggered QCS.}
Krco and Paul have proposed the measurement triggered QCS protocol for multiparty clock synchronization\cite{Krco2002}. Qubits which have two energy levels are used, labeled as $|0\rangle$ and $|1\rangle$, with energy $E_0$ and $E_1$ respectively, $\omega = (E_1-E_0)/\hbar$.  The protocol starts from an initial W state, which has the form
\begin{eqnarray}
|W(N)\rangle &=& \frac{1}{\sqrt{d}}(|10...0\rangle_{012...d}+|01...0\rangle_{012...d}\nonumber \\ && +\cdots+|00...1\rangle_{012...d}).
\end{eqnarray}
Each node contains a single qubit. The initial W state is an energy state, so that it remains invariant until measurement is made. At an agreed time, each node measures its qubit in the basis $\{ |+\rangle, |-\rangle \}$, where $|+\rangle = (|0\rangle+1\rangle)/\sqrt{2}$, $|-\rangle = (|0\rangle-|1\rangle)/\sqrt{2}$. Since the real time of each node is not the same, the measurements are made at different times. Take time $t_0$ as the standard time, we focus on the time of node $i$. After node $0$ measures its qubit, the measurement result is published, which will determine the results of node $i$'s measurements. The probabilities of getting $|+\rangle$ and $|-\rangle$ at node $i$ are
\begin{equation}
P(|\pm\rangle) = \frac{1}{2} \pm \frac{\cos(\omega (t_i-t_0))}{d+1},
\end{equation}
when the measurement result at node 0 is $|+\rangle$. $|-\rangle$ yields similar result. We see measurements at node $i$ allow the estimation of $t_i-t_0$. Similarly the estimation of other nodes' times are possible as well.

The trigger of this QCS protocol is measurement, which means that measurement starts the evolution of the state. Since qubits are distributed at different nodes, only local measurements are allowed.

\begin{figure}[tbp]

\centerline{\includegraphics[width= 6 cm]{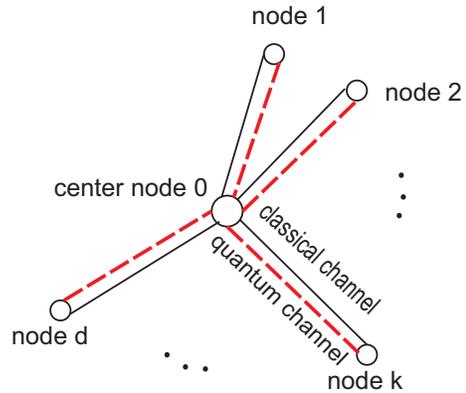}}
\caption{Framework of the operation triggered QCS network. Nodes from 1 to d are all connected with the center node 0. Quantum and classical channels are available for the communication of node 0 and node k, k = 1, ..., d. The solid line stands for the classical channel, and the dashed line stands for the quantum channel.}
\label{frm}
\end{figure}

\begin{figure*}[tbp]
\centerline{\includegraphics[width= 18 cm]{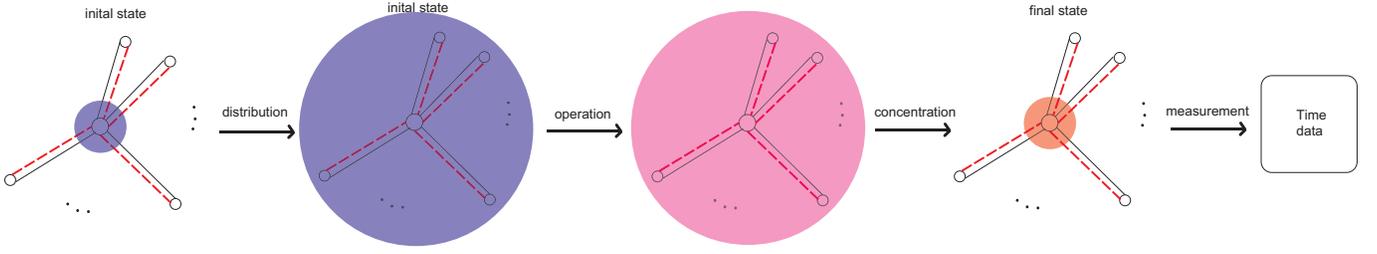}}
\caption{Framework of the clock synchronization network. Nodes from 1 to d are all connected with the center node 0. Quantum and classical channels are available for the communication of node 0 and node k, k = 1, ..., d. The solid line stands for the classical channel, and the dashed line stands for the quantum channel. The shaded area represents where the qubits are.}
\label{process}
\end{figure*}

\textit{Operation triggered QCS.}
The framework of our QCS protocol is shown in Fig.[\ref{frm}]. Nodes from 1 to d are all connected with the center node 0. Quantum and classical channels are available for the communication of node 0 and node k, k = 1, ..., d. The process of our clock synchronization scheme is shown in Fig.[\ref{process}]. The process can be divided into five stages. The first is the preparation stage, an initial state $|\Psi_i\rangle$ is prepared in node 0. We note that unlike the previous protocols, here the initial state $|\Psi_i\rangle$ is arbitrary. The second is the distribution stage, qubits are distributed to the $d$ nodes through the quantum channels, after which each node contains a certain number of qubits, and they on the whole form the initial state $|\Psi_i\rangle$. In these two stages, $|\Psi_i\rangle$ is an energy eigenstate such that it will remain invariant. The third is the triggering stage, the $d$ nodes make operations on the qubits they have owned, which will trigger the evolvement of the qubits at the nodes' local times. The evolvement will contain the information of the local times. After all the operations are done, the state stops evolving and becomes invariant. The fourth is the concentration stage, qubits are transmitted back to center node $0$ and after some operations we get the final state $|\Psi_f\rangle$, which contains the information of the times of the $d$ nodes. The fifth stage is the estimation stage, where measurements are made, after which the time differences of the $d$ nodes and the center node can be calculated. When the time information is obtained, it can be transmitted to the other nodes through the classical channels.

After the distribution stage, we denote the number of qubits node $k$ owns as $n_k$, $k = 0, 1, ..., d$, and $N = \sum_{k = 0}^d n_k$ is the total number of qubits used in the QCS protocol.
We suppose that the initial state $|\Psi_i\rangle$ can be written in the Fock basis as
\begin{equation}
|\Psi_i\rangle = \sum_a c_a |\psi_a\rangle = \sum_a c_a  |N^a_{0}\rangle_0  |N^a_1\rangle_1\cdots|N^a_d\rangle_d,
\end{equation}
where $c_a$ are coefficients and $|N_a^i\rangle_i$ are Fock states with $N_a^i$ qubits in energy level $E_1$, $i = 1, ..., d$. A qubit will evolve under the unitary operator $U = e^{-i\hat{H}t}$, where $\hat{H} = E_0|0\rangle\langle 0 |+E_1|1\rangle\langle 1|$. Since we have chosen energy state as the initial state, we have $\sum_{i=0}^d N_a^i = N$, for any $a$, where $NE_1$ is the energy of the initial state above the ground state. $|\Psi_i\rangle$ will remain invariant.

At the triggering stage, every node from $1$ to $d$ does $\sigma_x$ to its part of the total state, when its clock points at a beforehand agreed time $t_0$. We denote the real time of each node doing the $\sigma_x$ operation as $t_1, ... , t_d$. Since $\sigma_x = |0\rangle\langle 1|+|1\rangle\langle 0|$, it operates as a NOT gate. We suppose $t_1 < t_2 < \cdots < t_d < t_0$. We must make sure that $t_0$ is the biggest, which can be guaranteed by purposely delay the time of performing the $\sigma_x$ operation at node 0. The order of other times is not important since we can relabel the nodes so that the assume is correct, and because all the $d$ nodes are equivalent to each other, no practical change will happen.

Now the real situation is, all the nodes perform the $\sigma_x$ operation at respective times $t_1$, $t_2$, ..., $t_d$, $t_0$. Set $t_1 = 0$ as the starting time, when $t = 0^+$, which means the real time has just passed $t_1$, and node $1$ has just done the $\sigma_x$ operation, while the states of all other nodes remain the same, we have $|\psi_a(t=0^+)\rangle = |N^a_0\rangle|n_1-N^a_1\rangle|N^a_2\rangle\cdots|N^a_d\rangle$. We notice that after node $1$ has done the $\sigma_x$ operation, the total state isn't stationary any more, and starts to evolve with time. So in fact the $\sigma_x$ operation triggers the evolution of the total state. At time $t\rightarrow t_2$, which means that the time is infinitely close to $t_2$ but haven't reached it, we have
\begin{eqnarray}
|\psi_a(t\rightarrow t_2)\rangle &=& e^{-i N^a_0 \omega (t_2-t_1)}|N^a_0\rangle_0 \nonumber \\
 & & \otimes e^{-i(n_1-N^a_1)\omega (t_2-t_1)}|n_1-N^a_1\rangle_1 \nonumber \\
& & \otimes_{i=2}^d e^{-i N^a_i \omega (t_2-t_1)}|N^a_i\rangle_i \nonumber \\
&=& |N^a_0\rangle_0 \nonumber \\
& & \otimes e^{2i N^a_1\omega (t_2-t_1)}|n_1-N^a_1\rangle_1 \nonumber \\
& & \otimes_{i=2}^d |N^a_i\rangle_i,
\end{eqnarray}
with the same global phase omitted for all the states $|\psi_a\rangle$.

Similarly we have
\begin{eqnarray}
|\psi_a(t\rightarrow t_3)\rangle &=& |N^a_0\rangle_0 \nonumber \\
& & \otimes e^{2iN^a_1\omega (t_3-t_1)}|n_0-N^a_0\rangle_0 \nonumber \\
& & \otimes e^{2iN^a_2\omega (t_3-t_2)}|n_1-N^a_1\rangle_1 \nonumber \\
& & \otimes_{i=3}^d |N^a_i\rangle_i.
\end{eqnarray}

When $t>t_0$, the last $\sigma_x$ operation has been done at node $0$, and the total state stops to evolve, we then have
\begin{equation}
|\psi_a(t>t_0)\rangle = \otimes_{i=1}^d e^{2iN^a_i\omega (t_0-t_i)}|n_i-N^a_i\rangle_i
\end{equation}
and
\begin{equation}
|\Psi_i(t>t_0)\rangle = \sum_a c_a |\psi_a(t>t_0)\rangle.
\end{equation}

Then the triggering stage is over and we enter into the concentraion stage. After all the qubits are concentrated at the center node 0, we perform the $\sigma_x$ operation to every constituting qubit of $|\Psi_i(t>t_0)\rangle$ at node $0$ at the same time and obtain the final state
\begin{eqnarray}
|\Psi_f\rangle &=& \sum_a c_a \otimes_{k=0}^{d} e^{2i \omega N^a_k(t_0-t_k)}|N^a_k\rangle_k \nonumber \\
&=& e^{\sum_{k=1}^d 2i \omega \hat{n}_k(t_0-t_k)} |\Psi_i\rangle.
\end{eqnarray}

Write $\theta_k = t_k-t_0$, we have
\begin{eqnarray}
|\Psi_f\rangle = e^{-2 \omega \sum_{k=1}^{d} i \hat{n}_k \theta_k} |\Psi_i\rangle.
\end{eqnarray}
Up until now, by estimating $\theta_k$, $k = 1, 2,..., d$, which is a standard multiple phase estimation problem, we can get information about the time differences of clocks at different nodes.

\begin{figure}
\centerline{\includegraphics[width= 8 cm]{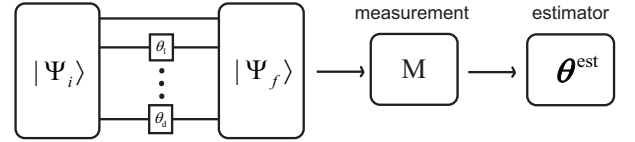}}
\caption{General scheme for quantum multiple phase estimation.}
\label{mulphase}
\end{figure}

A general protocol for multiple phase estimation in quantum metrology is shown in Fig.[\ref{mulphase}]. An initial state $|\Psi_i\rangle$ undergoes the process which depends on the phase vector $\bm{\theta}$, and evolves into the final state $|\Psi_f\rangle$. In our QCS protocol the phase vector is the time differences of the different nodes with the center node. Then we do measurements to the final states. The measurements are in association with the estimator we choose. From the results of the measurements, we will have an estimation $\bm{\theta}^{est}$ of the phase vector. It has been shown that simultaneous multiple phase estimation has some advantage over estimating the phases one by one, a realistic set-up is also proposed\cite{Humphreys2013}.

\textit{Two party QCS.}
We next study the special case of two party clock synchronization problem, where there are only two clocks to be synchronized, so $d = 1$ in this case. We show that the operation triggered QCS protocol will achieve the Heisenberg scale of the time difference estimation. For the two party case, we have
\begin{equation}
|\Psi_f\rangle = e^{-2 \omega  i \hat{n}_1 \theta_1} |\Psi_i\rangle,
\end{equation}
where $\theta_1 = t_1 - t_0$.

From the measurement results of $|\Psi_f\rangle$, we will have an estimate of $\theta$, denoted as $\theta^{est}$. The precision of the estimation of $\theta$ can be described by $\delta \theta$, where $\delta \theta = \sqrt{\langle  (\theta^{est} - \theta)^2 \rangle}$, the average is taken over all the possible measurement results. The Cram\'{e}r-Rao equality gives the ultimate precision of the estimation of $\theta$\cite{Paris2009}\cite{Helstrom1976},
\begin{equation}
\delta \theta \geq \frac{1}{\sqrt{\mu F_Q}},
\end{equation}
where $\mu$ is the number of measurements, $F_Q$ is called the quantum Fisher information, when the estimator is unbiased. We further note that for the single parameter case, the equality can be obtained, such that $\delta \theta = 1/\sqrt{\mu F_Q}$. Since we are only interested in the quantum enhancement of the precision, we will set $\mu = 1$, then $\delta \theta = 1/\sqrt{F_Q}$. For pure states, we have
\begin{equation}
F_Q = 4[\langle \partial_{\theta} \Psi_f|\partial_{\theta} \Psi_f\rangle + (\langle \partial_{\theta} \Psi_f|\Psi_f\rangle)^2].
\end{equation}
If we choose the initial state as the NOON state $|\Psi_i\rangle = \frac{1}{\sqrt{2}} (|0n\rangle + |n0\rangle)$, where each node has equal number of qubits and $n=N/2$, we will obtain the maximal quantum Fisher information $F_Q = 4\omega^2 n^2$. Then $\delta \theta = \frac{1}{2\omega n} = \frac{1}{\omega N}$. We see that it is the Heisenberg scale.

\textit{Average time}
In many cases, we don't need to know the times of all the clocks, what we are interested is the average time of the clocks, which is $\bar{t} = \frac{\sum_{i=k}^d t_k}{d}$. For example, the $d$ clocks may be clocks of satellites at various places, instead of synchronizing these clocks to set up a standard time, we may take the average time of the clocks as the standard time.  We show that when we want to synchronize the time of node 0 to $\bar{t}$, the operation triggered QCS strategy has an $O(\sqrt{d})$ advantage in precision over the best possible measurement triggered QCS strategy, and we will give explicitly the measurement operator and the estimator. We calculate the standard deviation of $\bar{\theta}$, where $\bar{\theta} = \bar{t} - t_0$.

For the operation triggered QCS, we select the initial state as
\begin{equation}
|\Psi_i\rangle = \frac{1}{\sqrt{2}}(|n_00\cdots 0 \rangle + |0 n\cdots n \rangle),
\end{equation}
where $n_0 = dn$, so that it is an energy state and will not evolve. Since $n_0+dn = N$, we have $n = \frac{N}{2d}$.

The final state is
\begin{eqnarray}
|\Psi_f\rangle &=& \frac{1}{\sqrt{2}}(|n_0 0\cdots 0 \rangle + e^{-2i\omega\sum_{k=1}^d n \theta_k} |0 n\cdots n\rangle )\nonumber \\
&=& \frac{1}{\sqrt{2}}( |n_0 0\cdots 0 \rangle + e^{-2i\omega n d\bar{\theta}} |0 n \cdots n\rangle).
\end{eqnarray}

The quantum Fisher information is
\begin{equation}
F_Q = (2\omega d n)^2.
\end{equation}
So we have $\delta \bar{\theta}_{opt} = 1/\sqrt{F_Q} = \frac{1}{2\omega d n } = \frac{1}{N\omega}$, where ``opt'' means operation triggered QCS.

For the measurement triggered QCS, we consider the best possible precision for the estimation of $\bar{\theta}$. Since the trigger is the measurement, only local measurement is allowed. There is no efficient way to directly make an estimation of $\bar{\theta}$, we can only estimate $\theta_k( k = 1, ..., d)$ one by one, and make an estimation of $\bar{\theta}$ from the equation $\bar{\theta} = (\sum_{k=1}^d \theta_k)/d$.

After the instant of node $0$'s measurement, the stationary initial state of all the nodes $|\Psi_i\rangle$, will collapse to a state $|\Phi_i\rangle$, which is not an energy eigenstate, so it begins to evolve.
We focus on node $1$, and see what is the best possible precision of $\theta_1$'s estimation. Suppose the collapsed state can be written as
\begin{equation}
|\Phi_i\rangle = \sum_{k=0}^{n_1} c_k |k\rangle_1 |\phi_k\rangle,
\end{equation}
where $|\phi_i\rangle$ stands for the state of all the other nodes, $|k\rangle_1$ is Fock state of node $1$ and $\sum_{k=0}^{n_1} |c_k|^2 = 1$.

After time $\theta_1$, it evolves to
\begin{eqnarray}
|\Phi(\theta_1)\rangle &=& exp(-i\theta_1 \sum_{j=0}^d\hat{H_j})\sum_{k=0}^{n_1} c_k |k\rangle_1 |\phi_k\rangle. \nonumber \\
   &=& \sum_{k=0}^{n_1} c_k e^{-ik\omega \theta_1} |k\rangle_1 |\phi_k'\rangle,
\end{eqnarray}
where an global phase omitted, $H_j$ is the evolution operator for node $j$, and $|\phi_k'\rangle = exp(-i\theta_1 \sum_{j\neq 1}\hat{H_j})|\phi_k\rangle$.

The optimal state to achieve the highest estimation precision of $\theta_1$ is
\begin{equation}
|\Phi_{opt}(\theta_1)\rangle = \frac{1}{\sqrt{2}}(|0\rangle_1 + e^{-in_1\omega \theta_1}|n_1\rangle_1)\otimes |\phi\rangle,
\end{equation}

where $|\phi\rangle$ is an arbitrary pure state of all the other nodes except node 1. In this way, the states that node 1 owns is a pure state $\frac{1}{\sqrt{2}}(|0\rangle_1 + e^{-in_1\omega \theta_1}|n_1\rangle_1)$. Node 1 can make local measurements of this state to estimate $\theta_1$ and the Fisher information is $F_Q = n_1^2\omega^2$. So $(\delta \theta_1)^2 =  \frac{1}{n_1^2\omega^2}$. Similarly we have $(\delta \theta_k)^2 =  \frac{1}{n_k^2\omega^2}$ for $k= 2, ..., d$.

Since $\bar{\theta} = (\sum_{k = 1}^d \theta_k)/d$, we have
\begin{equation}
\delta \bar{\theta} = \frac{\sqrt{\sum_{k=1}^d (\delta \theta_k)^2}}{d}.
\end{equation}

In order to give an lower limit of $\delta \bar{\theta}$, we further suppose that all qubits are distributed to node 1 to $d$, and node $0$ has no qubits, in this case when $n_k = N/d, k =1, 2, ..., d$, $\delta \bar{\theta}$ reaches the minimum
\begin{equation}
\delta \bar{\theta} > \frac{\sqrt{d\frac{d^2}{N^2\omega^2}}}{d} = \frac{\sqrt{d}}{N\omega}.
\end{equation}

We should note that the bound can not be reached, since the optimal state for measuring $\theta_1$ is not the optimal one for measuring other time differences, and at least some qubits out of the total $N$ qubits must be distributed to node $0$. We denote this bound as $\delta \bar{\theta}_{mea}$, where ``mea'' means measurement triggered QCS. Compare $\delta \bar{\theta}_{opt}$ and $\delta \bar{\theta}_{mea}$, we see that the operation triggered QCS has an $O(\sqrt{d})$ advantage in $\bar{\theta}$'s estimation over measurement triggered QCS. In fact, the bound $\delta \bar{\theta}_{mea}$ is quite loose. The existing measurement triggered QCS protocol provides much worse precision. For the work of \cite{Ren2012}, the protocol using the simultaneous bipartite entanglement gives the best precision
\begin{equation}
(\delta \theta_k)^2 = \frac{1}{4\omega^2}(\frac{N}{N-1})^2,
\end{equation}
if $N$ is big enough, we have $(\delta \theta_k)^2 = \frac{1}{4}$, which leads to $\delta \bar{\theta} = \frac{1}{2\omega\sqrt{d}}$. This protocol does not have the $\frac{1}{N}$ scale.

\textit{Conclusion.}
Clock synchronization is necessary in many modern fields of science and technologies. Quantum clock synchronization(QCS) uses the power of entanglement to synchronize remote clocks. The advantage of QCS is that the protocol is independent of the participants' relative locations and the properties of the intervening media, and the process of distributing entanglement is adiabatic. Existing QCS protocols use measurement as the trigger of starting the state's evolution. However, measurement will break the entanglement and only local measurement is allowed as a result. We propose a novel idea of operation triggered QCS, where we use unitary operation to trigger the evolution of the state. We turn the QCS problem to a standard quantum metrology problem, which is heavily studied and many knowledge can be borrowed. The operation triggered QCS makes general measurements of the total state possible. We show that the protocol achieves the Heisenberg scale in the two party case, which breaks the SQL limit. Also we show that our protocol is extremely efficient in estimating the average time. Operation triggered QCS has an $O(\sqrt{d})$ advantage of the precision of the avearge time's estimation over the best possible measurement triggered QCS protocol. The precision is also Heisenberg scale. In conclusion, we have proposed a novel QCS scheme, which has the Heisenberg scale of precision and very efficient in estimating the average time.

This work is supported by ``973" program (2010CB922904), NSFC and
grants from Chinese Academy of Sciences.

\end{document}